\documentclass[aps,prl,reprint,groupedaddress]{revtex4-2}
\usepackage[normalem]{ulem}
\usepackage{amsmath}
\usepackage[dvipsnames]{xcolor}
\usepackage{amssymb}
\usepackage{graphicx}
\usepackage{txfonts}

\usepackage{hyperref}
\hypersetup{
  colorlinks = true, % Colours links instead of ugly boxes
  urlcolor   = blue, % Colour for external hyperlinks
  linkcolor  = blue, % Colour of internal links
  citecolor  = blue  % Colour of citations
}



\begin{document}

\title{Momentum Flow Mechanisms and Color-Lorentz Forces on Quarks in the Nucleon}
\author{Xiangdong Ji}
\email{xji@umd.edu}

\author{Chen Yang}
\email{cyang127@umd.edu}

\affiliation{Maryland Center for Fundamental Physics, Department of Physics, University of Maryland, College Park, MD 20742, U.S.A.}
\date{\today}

\begin{abstract}

Momentum conservation in the nucleon is examined in terms of continuous flow of the momentum current density (or in short, momentum flow), which receives contributions from both kinetic motion and interacting forces involving quarks and gluons. While quarks conduct momentum flow through their kinetic motion and the gluon scalar (anomaly) contributes via pure interactions, the gluon stress tensor has both effects. The quarks momentum flow encodes the information of the color-Lorentz force density on them, and the momentum conservation allows to trace its origin to the gluon tensor and anomaly (a ``negative pressure'' potential). From the state-of-the-art lattice calculations and experimental fits on the form factors of the QCD energy-momentum tensor, we exhibit pictures of the momentum flow and the color-Lorentz forces on the quarks in the nucleon. In particular, the anomaly contributes a critical attractive force with a strength similar to that of a heavy-quark confinement potential. 

\end{abstract}
\maketitle

\section{Introduction}

Translational symmetry of the physical space leads to the momentum conservation law for isolated systems~\cite{Itzykson:1980rh,Noether1918}. By introducing the momentum density $k^i(\vec{r}, t)$ and momentum current density (MCD)
$T^{ij} (\vec{r}, t)$ ($i,j=1,2,3$ are spatial indices) representing the $i$-th direction flow of the momentum component $j$, one can examine the conservation law through the continuity equation,
\begin{equation}\label{eq:Continuity}
    \frac{\partial k^j(\vec{r}, t)} {\partial t} + \nabla_i T^{ij}(\vec{r}, t) = 0 \ ,  
\end{equation}
where $\nabla_i$ is the spatial differential operator. From the above, the total momentum $\vec{K} = \int {\rm d}^3\vec{r}~\vec{k}(\vec{r},t)$ is simply a constant of motion. The advantage of the continuity equation is that it depicts a dynamical picture of momentum conservation through continuously flowing its density over space.

The momentum density and MCD in strong-interaction systems are parts of the energy-momentum tensor (EMT) of quantum chromodynamics (QCD), $T^{\mu\nu}$ $(\mu,\nu=0,i)$, with $k^i=T^{0i}$~\cite{Collins:1976yq,Jaffe:1989jz,Ji:1994av,Ji:1995sv}. Their matrix 
elements in the hadrons like the proton and neutron, also referred to as the EMT form factors, have been found experimentally measurable in deep-exclusive processes such as deeply virtual Compton scattering or meson production through sum rules of generalized parton distributions (for a selected references, see~\cite{Ji:1996ek,Ji:1996nm,Muller:1994ses,Radyushkin:1996ru,Collins:1996fb,Ji:1998pc,Diehl:2003ny,Belitsky:2005qn,Burkert:2018bqq,Duran:2022xag,GlueX:2023pev,Guidal:2002kt,Boussarie:2016qop,Qiu:2023mrm,Guo:2021ibg,Mamo:2022eui,Guo:2025jiz}). They can also be studied theoretically through the lattice QCD calculations~\cite{Gockeler:2003jfa,Hagler:2003jd,LHPC:2007blg,Shanahan:2018nnv,Hackett:2023rif,Wang:2024lrm,Holligan:2020osy}. 
These form factors have provided a novel channel to explore the mass and spin structures of the nucleon~\cite{Ji:1994av,Ji:1995sv,Ji:1996ek}, as well as the momentum flow mechanism. 

In the last few years, we have seen an increasing interest in the so-called mechanical properties of the nucleon~\cite{Polyakov:2002wz,Polyakov:2002yz,Goeke:2007fp,Burkert:2018bqq,Polyakov:2018zvc,Lorce:2018egm,Kumericki:2019ddg,Liu:2021gco,Ji:2021mfb,Fu:2022rkn,Ji:2022exr,Liu:2023cse,GarciaMartin-Caro:2023klo,Czarnecki:2023yqd,Freese:2024rkr,Lorce:2025oot}. It has been suggested that the nucleon may be considered as a continuous medium in static equilibrium and the $T^{ij}$ form factor $C/D$~\cite{kobzarev1962gravitational,Pagels:1966zza,Ji:1996ek} can be interpreted as the mechanical stress tensor of pressure and shear force distributions~\cite{Polyakov:2002wz,Polyakov:2002yz}. The ``pressure'' distribution or the trace $T^{ii}(\vec{r})$, satisfying the following well-known relation derived from the continuity equation $\partial_i T^{ij}(\vec{r})=0$, 
\begin{equation}
    \int T^{ii}(\vec{r})~{\rm d}^3\vec{r}=0 \ , \label{eq:von-Laue}
\end{equation}
has been regarded as a mechanical stability condition~\cite{Goeke:2007fp} (the von Laue condition~\cite{Laue:1911lrk}). It has been interpreted as the balance of 2D surface forces inside the nucleon, where the positive pressure leads to the repulsive force while the negative pressure is attractive~\cite{Goeke:2007fp}. 

We believe, however, the above interpretation is subject to debate. {First of all, although all systems have MCD, but not all MCDs are simply pressure, as in the cases of a moving fluid and a beam of laser. Exactly when MCD becomes pressure requires careful examination which has not been carefully discussed in the context of QCD in the literature. Second, even when a part of or the whole MCD can be regarded as pressure, it does not readily mean the surface force between different parts of the system, as in ideal gases. Only in very special cases, the pressure corresponds to a surface force. In the case of nucleon,} the overall pattern of the momentum flow, described by the form factor $C/D$, may not have any overarching mechanical significance {as a pressure or force between different parts}. This is because the total MCD receives ambiguities from the superpotentials~\cite{BELINFANTE1940449,rosenfeld1940energy,Blaschke:2016ohs,Itzykson:1980rh} when the interactions are present. In terms of the momentum continuity, the specific form of $C$ is not that significant: any transformation, $C\rightarrow C+\Delta C $, makes the momentum conservation unchanged. It is also well known that Noether's theorem does not generate a unique conserved current. Therefore, there is no way to know {\it a priori} which version has more mechanical significance than the other. As far as we know, the only way to favor a special MCD is through its coupling to new interactions beyond QCD. Here the option is clearly gravity. However, the physical effect of this specific gravity-coupled MCD is limited to generating the spacetime metric perturbation and tensor monopole moment~\cite{Ji:2021mfb}.

In this paper, we seek to clarify the physics of MCD by studying the mechanisms of the momentum flow and related forces through its individual components $T^{ij}\equiv \sum_{\alpha} T^{ij}_{\alpha}$, following an EMT decomposition proposed in Ref.~\cite{Ji:1995sv}. There are in general two types of contributions for the momentum transport: kinetic motion of particles and mechanical forces. While the quark contribution is purely kinematical, the gluon contribution contains both physical gluonic radiations and static Coulomb interactions. The most intriguing aspect is the interaction contribution from the gluon scalar through the trace anomaly. This contribution is an attractive potential MCD (sometimes called ``negative pressure'') due to a change in the QCD vacuum in the presence of valence quarks. 

Only the momentum continuity equation, 
\begin{equation}
   \sum_{\alpha} \partial_i T^{ij}_{\alpha} =0 \ ,
\end{equation}
can be seen as the local balance of force densities through the change of local momentum from particles and interactions. We can identify forces on the quarks through the divergence of their kinetic MCD, which has a direct physical meaning as {\it color-Lorentz force}. We find the attractive force on quarks with a strongly confining component of the trace anomaly. The average strength of this anomaly force is approximately 1 GeV/fm, similar to the well-known QCD string tension~\cite{Sun:2020pda,Brambilla:2021wqs}. In this paper,  we consider that the light quarks are massless to simplify the discussion.

\section{QCD Mechanisms for the Momentum Flow in the Nucleon}

The QCD EMT and its renormalization has been worked out previously $T^{\mu\nu} = \bar T^{\mu\nu} + \hat T^{\mu\nu}$, where both the trace $\hat{T}^{\mu\nu}$ and traceless $\bar{T}^{\mu\nu}= \bar{T}^{\mu\nu}_q+\bar{T}^{\mu\nu}_g$ parts are scale and scheme-independent~\cite{Collins:1976yq,Ji:1994av,Ji:1995sv}. It is gauge-invariant and symmetric in $\mu\nu$, and is believed to be the gravitational charge. Specializing to the spatial components of the EMT, one obtains three contributions to the MCD,
\begin{equation}
    T^{ij}(\vec{r}) = \bar{T}^{ij}_q(\vec{r},\mu) + \bar{T}^{ij}_g(\vec{r},\mu) + \hat T^{ij}_a(\vec{r})  \ ,
\end{equation}
where $\mu$ is the renormalization scale, $\bar{T}^{ij}_q(\vec{r},\mu)$, $\bar{T}^{ij}_g(\vec{r},\mu)$, and $\hat{T}^{ij}_a(\vec{r})$ represent quark kinetic, gluon tensor and gluon scalar (trace anomaly) contributions to the total QCD momentum flow, respectively. 

The quark kinetic MCD $\bar{T}_{q}^{ij}(\vec{r},\mu)=(1/2)\bar{\psi}i\mathcal{D}^{(i}\gamma^{j)}\psi(\vec{r})$ represents the transport of momentum through the kinetic motion of quarks. This can be easily seen through the kinetic MCD of particles in the classical and non-relativistic limits, 
\begin{equation}
    T^{ij}_{K}(\vec{r}, t) = \sum_a p^i_a(t)v^j_a(t) \delta^{(3)}(\vec{r}-\vec{r}_a(t)) \ , \label{eq:EMT-Particle} 
\end{equation}
where $\vec{p}_a(t) =m_a\vec{v}_a(t) $, $m_a$ and $\vec{v}_a(t)$ are the momentum, mass and velocity of the $a$-th particle, respectively, and $\vec{r}$ represents the coordinate space distribution. By tracing $i,j$, we find that $T^{ii}_{K}(\vec{r})$ is proportional to the kinetic energy density $\varepsilon_K(\vec{r})$. Similarly, tracing the relativistic quark kinetic MCD also gives rise to the quark kinetic energy density distribution $\bar{T}^{ii}_{q}(\vec{r})=\bar{T}^{00}_{q}(\vec{r})\equiv \varepsilon_q(\vec{r})$ in the coordinate space. 

The gluon tensor MCD $\bar{T}_{g}^{ij}(\vec{r},\mu) =-(1/4)\delta^{ij}F^{2}-F^{i\alpha}F_{\ \alpha}^{j}(\vec{r})$ is the standard ``stress tensor'' of gauge fields as in the classical electromagnetism. While in the non-relativistic limit, this MCD conducts momentum flow mainly through the static Coulomb interactions as in the hydrogen atom, the quark dynamics in the nucleon is ultra relativistic and can generate the gluon radiation as an integral part of the system. These radiative gluons can carry the momentum flow through its kinetic motion like quarks. However, there is no frame- and gauge-independent separation of the radiations from the static gluonic interactions. Therefore, the gluonic MCD flows the momentum through a mix of kinetic motion and interacting forces. Similar to the quark case, by tracing $i,j$ indices, one obtains the gluonic energy density $\bar{T}^{ii}_g(\vec{r})=\bar{T}^{00}_g(\vec{r})\equiv \varepsilon_g(\vec{r})$. 

The anomaly MCD $\hat{T}_{a}^{ij}(\vec{r})=-(1/4)\delta^{ij}(\beta(g)/2g)F^{2}(\vec{r})$ is negative definite, and contributes to the total without a corresponding momentum density. Therefore, it reflects a pure interacting effect, and is also the negative of the anomalous energy density distribution $\hat{T}_{a}^{ij}(\vec{r}) \equiv -\varepsilon_a(\vec{r})\delta^{ij}$~\cite{Ji:2021qgo}. This MCD is essentially a negative pressure potential, reflecting that quarks ``sweep'' out the true QCD vacuum and lower the expectation of gluon condensate, as in the instanton liquid model~\cite{Zahed:2021fxk}, in line with the bag model phenomenology~\cite{Chodos:1974je,Chodos:1974pn,Johnson:1975zp,Thomas:1982kv}.

With the above decomposition, the physical meaning of the momentum flow balance in Eq. (\ref{eq:von-Laue}) reduces to,
\begin{equation}
    \int \langle T^{ii}(\vec{r})\rangle~{\rm d}^3\vec{r}=\int\left[\varepsilon_q(\vec{r})+\varepsilon_g(\vec{r})-3\varepsilon_a(\vec{r})\right]{\rm d}^3\vec{r}=0 \ , \label{eq:Virial}
\end{equation}
where $\langle T^{ii}\rangle(\vec{r}) =\varepsilon_q(\vec{r})+\varepsilon_g(\vec{r})-3\varepsilon_a(\vec{r})$ reflects the physical content of what has been identified as the ``pressure'' distribution in the literature~\cite{Polyakov:2002yz}. This equation is the same as the virial relation between the scalar and tensor energy contributions to the nucleon~\cite{Ji:1994av,Ji:1995sv,Lorce:2025oot}. 

The nucleon MCD currents in the coordinate space $\vec{r}$ can be related to the gravitational form factors (GFFs) $A_{q,g}(q^2)$, $B_{q,g}(q^2)$ and $C_{q,g}(q^2)$, defined as the matrix elements of the QCD EMT in the nucleon momentum states~\cite{kobzarev1962gravitational,Pagels:1966zza,Ji:1996ek}, 
\begin{align}
    \left\langle P^{\prime}\left|\bar{T}_{i}^{\mu\nu}\right|P\right\rangle &=\bar{U}(P^{\prime})\left[A_{i}(q^2)\gamma^{(\mu}\bar{P}^{\nu)}+B_{i}(q^2)\frac{\bar{P}^{(\mu}i\sigma^{\nu)\alpha}q_{\alpha}}{2M}\right. \nonumber \\
    +&C_{i}\left.(q^2)\frac{q^{\mu}q^{\nu}-q^{2}g^{\mu\nu}}{M}-\frac{M}{4}G_{s,i}(q^2)g^{\mu\nu}\right]U(P) \ ,\\ 
    \left\langle P^{\prime}\left|\hat{T}_{a}^{\mu\nu}\right|P\right\rangle &=\bar{U}(P^{\prime})\left[\frac{M}{4} G_{s}(q^2)g^{\mu\nu}\right]U(P) \ ,
\end{align}
where $i=q,g$ represent quarks and gluons, $M$ is the proton mass, $q^{\mu}=P^{\prime\mu}-P^{\mu}$ is the four-momentum transfer between the initial and final states with momenta $P^{\mu}$ and $P^{\prime\mu}$, $\bar{P}^{\mu}\equiv(P^{\prime\mu}+P^{\mu})/2$ is the average momentum. $G_s(q^2)=G_{s,q}(q^2)+G_{s,g}(q^2)$ is the scalar form factor, with $ G_{s,i}(q^2)$ arising from the non-conserving contributions~\cite{Ji:2021mtz},
\begin{equation}
    G_{s,i}(q^2)=A_i(q^2)+B_i(q^2)\frac{q^{2}}{4M^{2}}-C_i(q^2)\frac{3q^{2}}{M^{2}} \ .
\end{equation}
For simplicity, we will henceforth denote $\langle T^{ij} \rangle(q) \equiv \langle P^{\prime}|T^{ij}|P\rangle/2E_P$ and ignore the scale $\mu$ dependence. 

To obtain coordinate space $\vec{r}$-distributions, we consider a limit in which the nucleon is heavy so that the system recoils are negligible, $|q^2|\ll M^2$~\cite{Jaffe:2020ebz}. In practice, because of the nucleon mass is not super heavy, it is impossible to define strict classical-like 3D proton densities~\cite{Yennie:1957skg,Burkardt:2002hr,Miller:2018ybm,Jaffe:2020ebz,Epelbaum:2022fjc,Freese:2021czn,Lorce:2018egm}, and we choose the original Breit frame matrix elements~\cite{Sachs:1962zzc,Polyakov:2018zvc} and perform the Fourier transformation to get the spatial distributions,
\begin{align}
    \langle \bar{T}_{i}^{ij} \rangle (\vec{r}) &= \frac{M}{4}G_{s,i}(r)\delta^{ij}-\frac{1}{M}(\nabla^i \nabla^j -\delta^{ij}\nabla^2)C_{i}(r)\ , \label{eq:qgMCD}\\
    \langle \hat{T}_{a}^{ij}\rangle (\vec{r})&=-\frac{M}{4}G_{s}(r)\delta^{ij} \equiv p_a(\vec{r}) \delta^{ij}\ . \label{eq:aMCD}
\end{align}
where $G_s(r)\equiv\int\frac{{\rm d}^3 \vec{q}}{(2\pi)^3}e^{-i\vec{q}\cdot\vec{r}}G_s(-\vec{q}^2)$, and similarly for $C(r)$. Clearly, the quark and gluon momentum flows are in both radial and angular directions, which indicate that the average quark motion is analogous to an elliptic flow in fluid. Meanwhile, the position-dependent anomaly MCD, being negative, shows that the momentum flows towards the ``center''. It represents a ``negative pressure'' potential $p_a(\vec{r})=-\varepsilon_a(\vec{r})$ from the gluon scalar field expectation~\cite{Ji:2021qgo}) integrating to $-M/4$. By summing up these three components, the total MCD, 
\begin{equation}
    \sum_{\alpha}\langle T_{\alpha}^{ij} \rangle (\vec{r}) = -\frac{1}{M}(\nabla^i \nabla^j -\delta^{ij}\nabla^2)C(r) \ ,
\end{equation}
is expressed using only the $C/D$ form factor, which maybe considered as the MCD sum rule.

\begin{figure}[ht]
    \centering
    \includegraphics[width=0.82\linewidth]{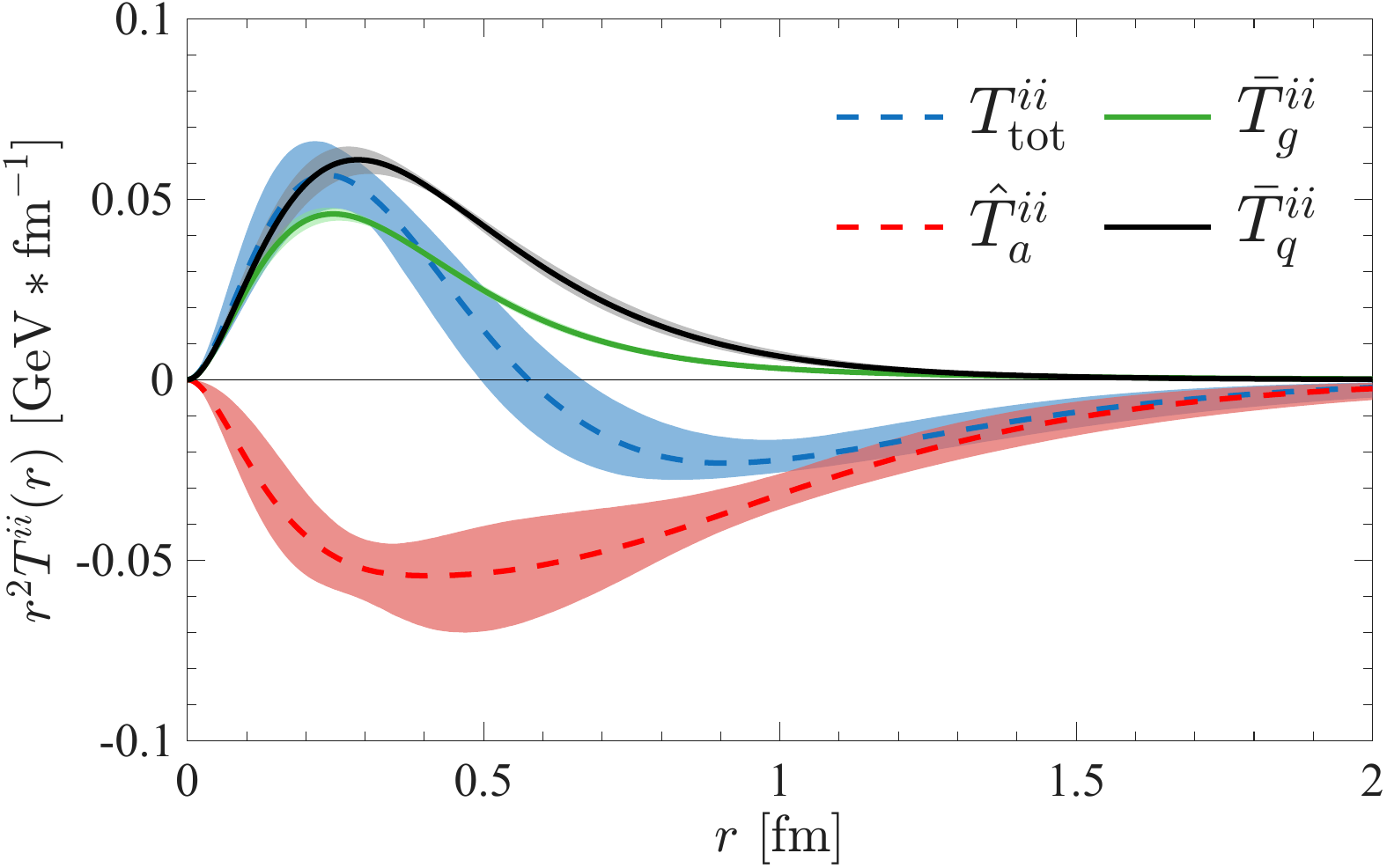}
    \caption{Trace of the momentum current distribution in proton and its decomposition into three components in Eqs.~(\ref{eq:Trace-qg}) and (\ref{eq:Trace-a}), illustrated using the latest phenomenological fits to both lattice QCD calculations and experimental data~\cite{Guo:2025jiz}. The total current (blue dashed) satisfying the virial theorem in Eq.~(\ref{eq:Virial}) is decomposed into the positive quark kinetic (black solid) and gluon tensor (green solid) contributions and the negative trace anomaly (red dashed) contribution. Uncertainties of 90\% confidence interval are shown as shaded areas.}
    \label{fig:Proton_Trace}
\end{figure}

We use the most recent phenomenological analysis on GFFs~\cite{Guo:2025jiz} to  exhibit the state-of-the-art knowledge on the momentum flow in the nucleon: contributions from different components and the total. This global analysis utilizes the Bayesian inference to fit various lattice QCD calculations and experimental data~\cite{Guo:2025jiz}. Specifically, the GFFs $F_{i}(q^2)=\{ A_{q,g}(q^2),C_{q,g}(q^2)\}$ are parametrized in the momentum space as $F_{i}(q^2)=F_{i0}(1-q^2/M_{F_{i}}^{2})^{-\alpha}$ with $\alpha=2$ for $A_{q,g}$, $\alpha=3$ for $C_{q,g}$. Notably, the $B$-form factor is relatively small and has been neglected. Due to the correlations among the parameters $\{F_{i0},M_{F_{i}}\}$, we utilize $2\times 10^4$ parameter sets obtained from the Monte-Carlo Markov Chain (MCMC) sampling method~\cite{Guo:2025jiz}. 

Based on the above parametrization and fits, we plot the trace of different components of the QCD MCD in FIG.~\ref{fig:Proton_Trace},
\begin{align}
    \langle\bar{T}_{q,g}^{ii}\rangle(\vec{r})&=\frac{3}{4}MA_{q,g}(r)-\frac{1}{4}\frac{1}{M}\nabla^{2}C_{q,g}(r) \ , \label{eq:Trace-qg}\\
    \langle\hat{T}_{a}^{ii}\rangle(\vec{r})&=-\frac{3}{4}MA_{q+g}(r)+\frac{9}{4}\frac{1}{M}\nabla^{2}C_{q+g}(r) \ .\label{eq:Trace-a}
\end{align}
By propagating the MCMC samples through these formulas---which preserves the full parameter correlations---we show the medians as curves, and the uncertainties are estimated using 90\% confidence intervals, shown as the shaded areas.

The quark and gluon contributions, or equivalently the quark and gluon energy density distributions, are found to be positive definite and dominate the small-$r$ behavior of the total MCD trace $\langle T^{ii}\rangle(\vec{r})$ (blue dashed curve in FIG.~\ref{fig:Proton_Trace}). The positive gluon MCD indicates that the radiative gluons may dominate the contribution, in contrast to the hydrogen atom, the electromagnetic MCD is negative due to the dominant contribution of the static Coulomb force~\cite{Ji:2022exr}. The negative anomaly contribution $p_a(r)$ accounts for the total trace being negative at large $r$, which explains the sign of the so-called $D$-term~\cite{Goeke:2007fp}. Physically, the negative anomaly contribution at large $r$ is expected from bag model phenomenology in which the scalar field with a large radius is present to confine the motion of quarks~\cite{Bogolubov:1968zk,Chodos:1974je,Thomas:1981vc}. 

Note that the current data on the anomaly MCD has the largest uncertainty, which is dominated by the errors of $C/D$ form factor (see FIG. 3 of Ref.~\cite{Guo:2025jiz}). Therefore, enhancing the future data precision and reducing uncertainties on $C/D$ are essential for achieving a better quantitative understanding of the momentum flow mechanism.

\section{Color-Lorentz Forces on Quarks}

As discussed earlier, momentum flows can arise from interacting forces, but in general they are not the 2D surface forces directly, even though the momentum current $T^{ij}$ has the unit of pressure. Only for the systems with zero-range forces as in liquids and solids, can the interaction part of the MCD $T^{ij}$ be interpreted as internal 2D pressure/shear forces~\cite{landau2013fluid}. The color forces among quarks have a range comparable to the size of the nucleon and therefore, no part of the MCD can be interpreted as the 2D surface forces among quarks and gluons.

Nonetheless, we can learn about the internal forces inside the nucleon through the continuity equation of separate MCDs $T^{ij}_{q,g,a}$. For a classical system consisting of $N$ interacting particles with a two-body potential $V_{ab}(|\vec{r}_a-\vec{r}_b|)$, the local continuity equation of the kinetic EMT gives, 
\begin{equation}
    \frac{\partial T^{0j}_{K}(\vec{r},t)}{\partial t} + \partial_i T^{ij}_{K}(\vec{r},t) = \sum_a F^j_a(t) \delta^{(3)} \left(\vec{r}-\vec{r}_a(t)\right) \equiv {\cal F}^j(\vec{r})\ ,\label{eq:Force}
\end{equation}
where we have used Newton's second law and $F_a^j$ is the internal force acting on the $a$-th particle from the others. This equation yields the {\it force density} ${\cal F}^j(\vec{r})$ exerted on the particles. Similarly, in the example of a hydrogen atom, it is easy to show that the electron's kinetic MCD~\cite{Ji:2022exr,Freese:2024rkr} also satisfies, 
\begin{align}
    \partial_{i} \langle T^{ij}_{e-K} \rangle(\vec{r}) =-\frac{\partial V(\vec{r})}{\partial r^{j}}\left|\psi_e(\vec{r})\right|^{2} =F^{j}\left|\psi_e(\vec{r})\right|^{2} \equiv {\cal F}^j(\vec{r})\ , \label{eq:e-Force}
\end{align}
where $\psi_e(\vec{r})$ is the electron wave function, $V(\vec{r})$ is the electric potential, $\vec{F}(\vec{r})$ is the Coulomb force, and $\vec{\cal F}(\vec{r})$ is the force density (weighted by the electron probability distribution). 

Therefore, the force can be learned from the divergence of the kinetic MCD, or classically the momentum change of a moving particle (see also~\cite{Polyakov:2018exb,Won:2023zmf,Freese:2024rkr}). We define the force density on quarks as,
\begin{equation}
  {\cal F}_q^j (\vec{r}) \equiv \partial_\mu \langle \bar{T}^{\mu j}_q \rangle = \partial_i \langle \bar{T}^{ij}_q \rangle(\vec{r})
  = \frac{M}{4}\nabla^j G_{s,q}(r)\ ,
\end{equation}
which is weighted by the quark probability density implicitly. ${\cal F}_q^j$ is exactly the standard {\it color-Lorentz force} density, 
\begin{align}
    \partial_{\mu}\bar{T}_{q}^{\mu j}
    &=gF_{a}^{\mu j}\bar{\psi}\gamma_{\mu}t_{a}\psi \nonumber \\
    &=g\rho_{a}E_{a}^{j}+g\left(\vec{j}_{a}\times\vec{B}_{a}\right)^{j}\equiv {\cal F}^j_q
\end{align}
where $(\rho_a,\vec{j}_a)=\bar{\psi}\gamma^{\mu}t_{a}\psi$ is the color current, which is different from previous works~\cite{Burkardt:2008ps,Crawford:2024wzx,Liu:2025ypg} in the infinite momentum frame.

On the other hand, the continuity equation in Eq. (\ref{eq:Continuity}) allows us to trace the force on quarks to the divergences of the gluon tensor and anomaly MCDs, 
\begin{equation}
    \partial_i \langle \bar{T}^{ij}_q \rangle (\vec{r})=- \partial_i \langle \bar{T}^{ij}_g \rangle(\vec{r}) - \nabla^j p_a(\vec{r}) \equiv {\cal F}_{g}^j(\vec{r}) + {\cal F}_{a}^j(\vec{r}) \ ,
\end{equation}
where the force densities are exclusively along the radial direction due to the spherical symmetry of the static nucleon. The above equation is analogous to Euler's equation for a fluid~\cite{landau2013fluid}, where the anomaly contribution $-\nabla^i p_a(\vec{r})$ is similar to a ``pressure term'' (but is rather a scalar potential~\cite{Ji:2021qgo}), and the gluon tensor contributes as the electron-proton interaction in Eq.~(\ref{eq:e-Force}). 

However, we strongly warn against taking the comparison with a fluid too literally: while the above quantum mechanical average is universally correct as an Ehrenfest theorem, fluid dynamics requires further assumptions to turn the above into a dynamical equation for infrared degrees of freedom (fluid elements containing many particles in local thermal equilibrium) through averaging over the ultraviolet ones (scales much smaller than fluid elements), which is inapplicable for the nucleon~\cite{landau2013fluid}. 

\begin{figure}[ht]
    \centering
    \includegraphics[width=0.83\linewidth]{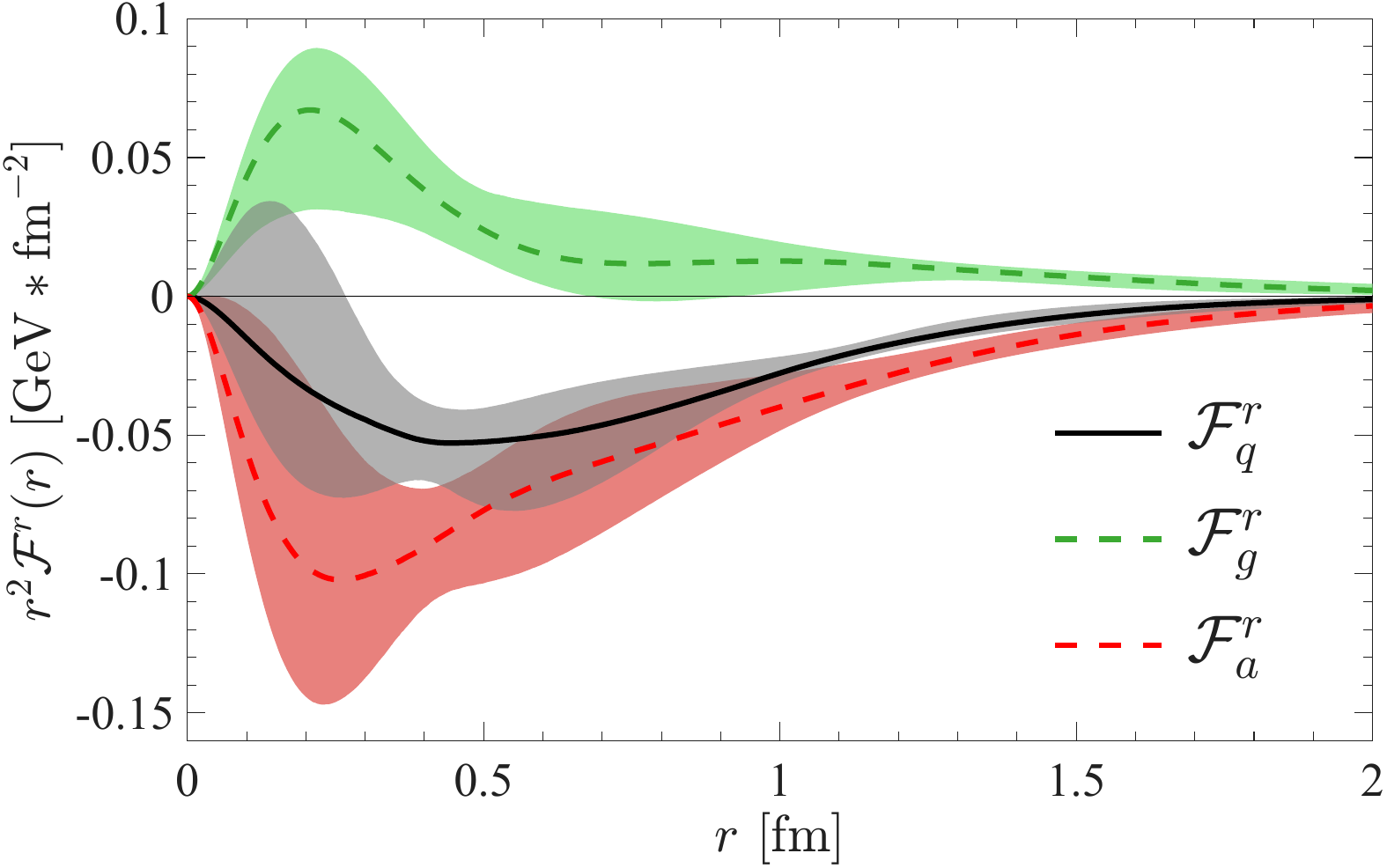}
    \caption{Force density distributions acting on quarks in the proton, visualized using the latest phenomenological fits to lattice QCD calculations and experimental data~\cite{Guo:2025jiz}: the large attractive force from the anomaly (red dashed) and the repulsive force from the gluon tensor (green dashed) combine to produce the total confining force (black solid) on quarks. The uncertainties are shown as shaded areas.}
    \label{fig:Forces}
\end{figure}

We again take the state-of-the-art global analysis utilizing various lattice QCD calculations and experimental data of the GFFs~\cite{Guo:2025jiz} to illustrate the force density distributions within the proton. We plot the radial components of the above force densities in FIG.~\ref{fig:Forces}. The uncertainties are determined similarly by propagating the MCMC samples of $\{F_{i0},M_{F_i}\}$~\cite{Guo:2025jiz} to the force densities through the above formulas.

Notably, due to the rather substantial uncertainties and the nucleon recoil effects, distributions at small $r$ may not reflect the realistic nucleon structures accurately, underscoring the urgency of improving data precision. Therefore, we focus on the results at relatively large $r$ region, and find that the total force density acting on quarks is attractive towards the ``center'' with negative values, which reveals the mechanical mechanism of the quark confinement. Tracing to individual components, we find the force from the gluon tensor is positive/repulsive, indicating perhaps the radiative gluon dominance, while the anomaly force is larger and attractive. 

Despite the large uncertainties and the recoil effects, we nevertheless calculate the average color-Lorentz force by integrating over $\vec{r}$. The average radial force from the anomaly is, 
\begin{equation}
    \bar{{\cal F}}_a \equiv \int {\rm d}^3 \vec{r} \ {\cal F}_a(\vec{r}) = -1.06^{+0.12}_{-0.11} \ {\rm GeV/fm} \ .
\end{equation}
This average anomaly force $\bar{{\cal F}}_a$ exhibits a magnitude similar to the confinement string tension~\cite{Sun:2020pda,Brambilla:2021wqs}, a concrete evidence that the anomaly plays an important role in the quark confinement~\cite{Rothe:1995hu,Liu:2021gco,Shuryak:2021yif,Liu:2024rdm}. Notably, one previous model estimation has found a value an order of magnitude smaller~\cite{Polyakov:2018exb}.

\section{Conclusion and Discussion}

In this paper, we have studied the underlying physical mechanism of each component of the QCD momentum flow. We find that quarks carry the momentum density through the ``collective'' motion both in the radial and angular directions. The gluon tensor flows momentum both as momentum carriers and as a mediator of forces among quarks, while the gluon scalar (trace anomaly) contributes through the pure interacting effect, resulting in a negative ``pressure potential'' due to a Casimir-like effect, the change of the QCD vacuum. The total MCD simply shows the momentum conservation through the balance of the in and out flows of the momentum density. 

We also find that the force effects can be studied through the divergences of the QCD MCD, which give the color-Lorentz force densities. Within the proton, the total force acting on quarks is found to be strongly confining. This force receives contributions from the gluon tensor and trace anomaly components, where the anomaly provides a large attractive force with an average magnitude of 1 GeV/fm, similar to the QCD string tension, indicating the origin of confinement. Notably, these force densities are related to the scalar form factor $G_s(q^2)$ rather than the $C/D$ form factor.

Finally, we want to comment that the physics of the MCD in the infinite momentum frame is simplified: only the $T^{++}$ component ($V^+=(V^0+V^3)/\sqrt{2}$) is leading and has a density interpretation (others are higher twists), which represents the momentum flow carried by unpolarized partons along the direction of proton motion. It is just the velocity (speed of light) multiplied by momentum density, similar to that of a beam of particles, as expected. In this case, form factor $C$ drops out, which together with other subleading MCD components does not have simple density interpretation.  Thus a force discussion in such a frame is beyond the scope of this paper.  

\section*{Acknowledgment}
We thank Adam Freese, Zijian Li, Keh-Fei Liu, Zein-Eddine Meziani, Yushan Su, Jinghong Yang, Feng Yuan, and Ismail Zahed for useful discussions. We also thank Yuxun Guo for providing the data for the parametrization of GFFs. XJ and CY are partially supported by Maryland Center for Fundamental Physics (MCFP). CY also acknowledges partial support by the U.S. Department of Energy, Office of Science, Office of Nuclear Physics under the umbrella of the Quark-Gluon Tomography (QGT) Topical Collaboration with Award DE-SC0023646.

% \bibliographystyle{apsrev4-1}
% \bibliography{ref}

%

\end{document}